# Waste Management Hackathon Providing New Ideas to Increase Citizen Awareness, Motivation and Engagement

Inna Sosunova
Software Engineering department
Lappeenranta–Lahti University of Technology LUT
Lappeenranta, Finland
inna.sosunova@lut.fi

Jari Porras
Software Engineering department
Lappeenranta–Lahti University of Technology LUT
Lappeenranta, Finland
jari.porras@lut.fi

Ekaterina Makarova
Digital Control Systems department
ITMO University
Saint-Petersborg, Russia
enrebelle@gmail.com

Andrei Rybin
Center of Information Optical Technology
University of Jyväskylä
Jyväskylä, Finland
andrei.rybin@gmail.com

*Abstract*—This paper describes the International Disruptive Information Solutions hackathon and one the winning solutions. The purpose of the hackathon was to promote the use of disruptive ICT technologies (e.g. IoT, Big data, AI, blockchain) in urban infrastructures to create innovative waste management solutions in a smart city context. 29 students enrolled into this hackathon and in the end 4 teams submitted their solutions to the challenges. The winning proposal EcoQ, an approach for plogging – collecting trashes while jogging, answered more than well to the presented challenge on waste management and engagement. The original idea was extended and partly refocused during an internship. As the outcome of the internship a mobile application for organizing and holding waste collection events was developed. This mobile application was shortly tested in a real environment and it provides a working citizen-centric platform, which enables anyone to arrange waste management events, and motivates other residents to participate in these activities.

## I. Introduction

A hackathon is an event where developers work together to solve a problem within a given timeframe [1] [2]. In this study we use hackathon as a tool for generating new solutions and confirming previously formulated concepts in the area of smart waste management in a smart city context. The International Hackathon in Disruptive Information Solutions [3] was arranged as a part of ENI CBC (European Neighborhood Instrument Cross-Border Cooperation) funded project CroBoDDIT - Cross-Border Dimensions of Disruptive Information Technologies [4]. The hackathon had three main themes 1) From data to services in smart city waste management, 2) Solutions for engaging citizens to waste management and 3) Robots and technology as waste collectors and waste handling help.

This paper describes the outcome of the theme 1 "From data to services in smart city waste management". In theme 1 the participants were encouraged to develop new smart waste management solutions in a smart city context. Participants were instructed to think of the solutions in a "technology-information-service" context, meaning that in order to develop a service, what information is needed for that and in order to use information where to get that. There was a possibility to work on smart garbage bin (SGB) level (low level) or city-level (high level).

## II. Hackathon realization and methodology

Hackathon was organized as a three-day event on Apr 23 – 25, 2021. The event was open to students from any university as it was arranged as an online event. Teams consisting of one to four members were allowed. The themes were announced in the opening session of the hackathon, so the teams had only 72 hours to complete their innovation and development work. Each team had a nominated mentor (PhD student on the topic area) to help them in their innovation tasks. By the end of the event the teams were instructed to submit their solutions to the hackathon platform (Devpost) along with a short video (3 min or less) in which they demonstrate their idea to a panel of judges. Full implementation was not required but the mockups and first solution prototypes needed to be demonstrated in the final presentation. Submissions ranged from proof-of-concept solutions to fully functional apps. As a prize, LUT University offered the opportunity for a short (2-4 weeks) virtual research internships for the hackathon winners (of different themes). The purpose of the internship was to develop the hackathon idea into a working product with the help of the mentor. All interns received certificates and references from their mentors.

## III. Results

The hackathon was attended by 29 participants. In the end 4 teams, «EcoQ», «Islanders», «Greenify» and «Erudite», submitted their solutions for evaluation. Topic «From data to services in smart city waste management» was won by team «EcoQ» with the idea of a mobile application for waste collection and sorting while jogging.

The EcoQ original idea includes two parts: mobile application for plogging and a smart garbage bin solution. Plogging is a combination of two words: jogging and Swedish phrase for pick up, 'plocka upp'. This is an activity that combines jogging with garbage collection. The main features of the original application were 1) a map, which indicates the nearest points of separate waste collection, places and routes for waste collection; 2) a system for scoring points for the amount of garbage handed over by each user; 3) image recognition for waste type verification; 4) QR code generator. Smart garbage bin solution consisted of GPS, ultrasonic fill level sensor, scales, QR-code scanner and lid actuator. The aim of the application is making garbage picking up and segregation an everyday habit. This solution was developed as part of hackathon topic 1, but was also relevant to topic 2 "Solutions for engaging citizens to waste management". While finalizing the concept and implementation of the application, we decided to focus on the citizens' engagement.

During the internship the original idea of Plogging as an individual task was developed into more collaborative movement with a goal to clean areas. Thus, during internship, based on the original EcoQ idea, a mobile app for organizing and implementing public events for solid waste collection was implemented. We formulated the following goals for the application: 1) to clean up the area where the event is taking place; 2) to motivate residents to collect garbage on their own after the event; 3) to motivate residents not to throw garbage in the forest and on the street. We also developed the application scenario, that includes 4 steps: 1) definition of the solid waste collection event; 2) preparations for the event; 3) carrying out a solid waste collection event; 4) completion and results of the event. As a result of internship the mobile application for organizing and implementing public events for solid waste collection was developed. Thus the application has two sides: one for event arrangements (Figure 1) and one for participation (Figure 2) following the original idea of plogging.

Using the developed application, it is possible to organize garbage collection events (Figure 1) in a selected area, involve interested participants, mark the most polluted areas and create tasks for participants, evaluate the results of events and the amount of collected garbage of each type. The scoring system and the table of participants, sorted by the number of points scored during the event, are additional motivation. In the future, it will also be possible to involve sponsors in events and give prizes from sponsors to the best participants. Participant of the garbage collection events may find in the list of events the most suitable one by location and time; at the event, choose a task (for example, collecting only a certain type of garbage), a form of participation (in a team or separately) and a team to join or create own team.

Thus, the created application solves the problem of motivation and engagement of participants through increased environmental awareness, reward-based motivation, and social motivation. It also focus not on abstract environmental issues, but on issues that directly affect the citizens. Participants can choose an event taking place in a specific location or organize it themselves and cleaning up the territory near their residence, school or university. This significantly increase the number of potential users. During the events, it is also planned to raise environmental awareness among interested participants. We also expect to create an eco-community, promote the sorting of more types of waste and eco-friendly lifestyle during the events.

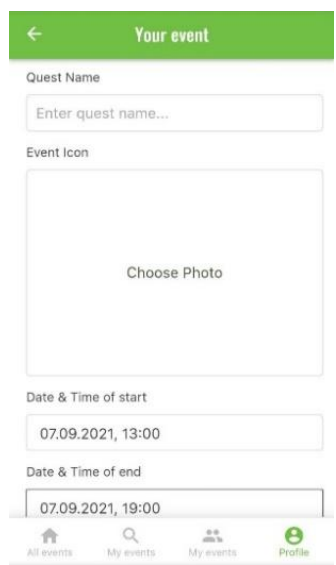
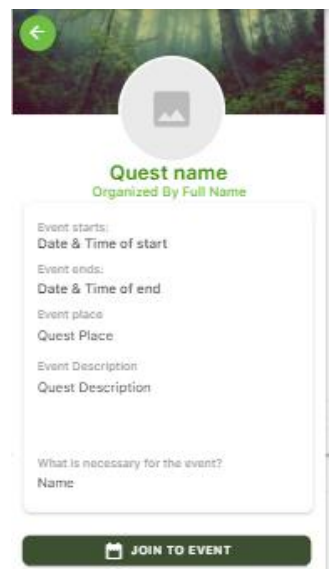

Fig. 1. EcoQ App: «Event organizer» role. Adding event data: name, icon, date, time

Fig. 2. EcoQ App: «Event participant» role. Event page

The EcoQ application was shortly tested in real conditions during a small test event with 5 participants. During the event, the application developers checked the operation of all functions, as well as the correctness of the generated data (the amount of garbage collected, the start time for completing tasks etc.). The following functions have been tested: creating quests, participating in quests, counting the collected garbage bags and generating a .csv-file with the data collected during the event.

## IV. CONCLUSION

In this study we described, discussed and analyzed Disruptive Information Solutions hackathon goals, methodology and results. Based on hackathon results we conclude that: 1) citizens engagement and motivation can be achieved through increased environmental awareness, reward-based motivation, or social motivation; 2) to increase citizens' ecological awareness, it is necessary to provide citizens with information about existing environmental problems, ways to solve them, and explain the importance and significance of the personal contribution of each. The outcomes of the hackathon will help to establish a new democratic and citizen-centric platform, which will enforce a transparent and efficient way of running municipal services, create vertical IoT applications, which will offer municipalities solutions to increase efficiency within their operations. In the future, we plan to hold a large garbage collection events using the developed application in Lappeenranta for students and staff of LUT University.

## ACKNOWLEDGMENT

This research has been carried out in the scope of the CroBoDDIT project (KS1592).